\documentclass[twocolumn]{article}%
\usepackage{epsfig,dcolumn}
\title{Phase Structures of Compact Stars in the Modified Quark-Meson Coupling Model}
\author{Chang-Qun Ma and Chun-Yuan Gao\footnote{Electronic address: gaocy@pku.edu.cn}\\
School of Physics, Peking University, Beijing 100871, China}
\begin{document}
\begin{titlepage}
\maketitle \noindent{\bf Abstract}

The K$^-$ condensation and quark deconfinement phase transitions are
investigated in the modified quark-meson coupling model. It is shown
that K$^-$ condensation is suppressed because of the quark
deconfinement when $B^{1/4}<$202.2MeV, where $B$ is the bag constant
for unpaired quark matter. With the equation of state (EOS) solved
self-consistently, we discuss the properties of compact stars. We
find that the EOS of pure hadron matter with condensed K$^-$ phase
should be ruled out by the redshift for star EXO0748-676, while EOS
containing unpaired quark matter phase with $B^{1/4}$ being about
180MeV could be consistent with this observation and the best
measured mass of star PSR 1913$+$16. We then probe into the change
of the phase structures in possible compact stars with deconfinment
phase as the central densities increase. But if the recent inferred
massive star among Terzan 5 with M$>$1.68M$_{\odot}$ is confirmed,
all the present EOSes with condensed phase and deconfined phase
would be ruled out and therefore these exotic phases are unlikely to
appear within neutron stars.

\end{titlepage}
Neutron stars are some of the densest objects in the universe since
their masses are of the order of 1.5 solar masses while their radii
are only of $\sim12$ km\cite{l}. Therefore, the density in the inner
core of a neutron star could be as large as several times nuclear
saturation density($\cong0.17$ fm$^{-3}$) and the appearance of new
phases other than normal nuclear matter is possible. Kaplan and
Nelson proposed the possibility of kaon condensation by a chiral
theory\cite{k}. While Bodmer\cite{b} and Witten\cite{w} suggested
that the strange quark matter phase, which was discussed by Itoh in
1970\cite{i}, might provide the absolutely stable form of the dense
matter. Following their work, many authors have devoted to studying
the kaon condensation and/or quark matter phase in neutron
stars\cite{gg}. In this letter, neutron star matter will be
investigated with novel EOSes and the possibilities of appearances
of exotic phases in neutron stars are going to be discussed.

Predictions by models with quark effects would be preferred to those
by models with only hadron degrees of freedom because neutron star
matter is extraordinarily dense. While at the moment QCD is not
realized in investigating neutron stars because of its
nonperterbative features, it is worthwhile studying the neutron star
with effective quark models. In 1988, Guichon proposed a novel
model\cite{g1}, the quark-meson coupling (QMC) model, where the
`quark effect' was incorporated. The model and its modified versions
give satisfactory description for saturation properties of nuclear
matter\cite{st} and reproduce the bulk properties of finite nuclei
well\cite{gsr}. Recently, Panda, Menezes and Provid$\hat{\rm e}$ncia
discussed the kaon condensation\cite{m} and deconfinement
phenomena\cite{p} in neutron star within the QMC model, where the
bag constant was fixed at its free-space value and the strange quark
was unaffected in the medium and set to its constant bare mass
value. But the QMC model predicts much smaller scalar and vector
potentials for the nucleon than that obtained in the well
established quantum hadrodynamics model. Jin and Jennings modified
the QMC model by introducing a density-dependent bag constant so
that large scalar and vector potentials are obtained without
affecting its abilities in other aspects\cite{j}. It is imaginable
that the $s$ quark mass should also be modified at the supernuclear
density in the core of neutron stars. So an additional pair of
hidden strange meson fields ($\sigma^*,\ \phi$), which had been
proved that they can account for the strongly attractive
$\Lambda\Lambda$ interaction observed in hypernuclei that cannot be
reproduced by ($\sigma,\ \omega,\ \rho$) mesons only\cite{sd}, are
included in the modified quark-meson coupling model (MQMC)\cite{ph}.
($\sigma^*,\ \phi$) couple only to the $s$ quark in the MQMC model
and only to the hyperons in the QHD model. The improved MQMC model
has been used to study kaon production in hot and dense hypernuclear
matter\cite{zg}.

In the present work, we shall extend the MQMC to investigate both
the K$^-$ condensation and quark deconfinement phase transitions in
compact stars at zero temperature. All of the three most possible
phases, i.e. the hadronic phase (HP) with strangeness-rich hyperons,
the condensation for negative charged kaon and the quark matter
phase are considered.

Both baryons and kaon meson are described by static spherical MIT
bags. Quarks are taken as explicit degrees of freedom, and are
coupled to the meson fields. The nonstrange ($u$ and $d$) quarks in
the baryons and kaons are coupled to the well known $\sigma$,
$\omega$ and $\rho$ meson fields while the strange quark in the
baryons and kaons is coupled to $\sigma^{*}$ and $\phi$ only,
because the former three pieces are built out of $u$-, $d$- quarks
or their anti-counterparts and the later two are composed of strange
quarks. Let the mean fields be denoted by $\sigma$, $\sigma^*$ for
the scalar meson fields, and $\omega_{0}$, $\phi_{0}$ and
$\rho_{03}$ for expectation valus of the timelike and the isospin
three-component of the vector and the vector-isovector meson fields.
In the mean field approximation the dirac equation for a quark field
of flavor $q\equiv(u,\ d,\ s)$ in the bag for the hadron species
$i\equiv({\rm p},\ {\rm n},\ \Lambda,\ \Sigma^{+},\ \Sigma^{0},\
\Sigma^{-},\ \Xi^{0},\ \Xi^{-},\ {\rm K}^-)$ is then given by
\begin{eqnarray}&&\left[{\rm i}\gamma\cdot\partial-m_
q+\left(g_{\sigma}^{q}\sigma-g_{\omega}^{q}\omega_{0}\gamma^{0}-g_{\rho}^{q}I_{3i}\rho_{03}\gamma^{0}\right)\right.\nonumber\\
&&\left.+\left(g_{\sigma^{*}}^{q}\sigma^{*}-g_{\phi}^{q}\phi_{0}\gamma^{0}\right)\right]\psi_{qi}(\vec{r},t)=0.\end{eqnarray}
The normalized ground state is solved as
\begin{equation}\psi_{qi}(\vec{r},t)=\mathcal{N}_{q}\exp\left(-{\rm i}\epsilon_{qi}t/R_{i}\right)\left(
                                                                               \begin{array}{c}
                                                                                 j_{0}\left(x_{qi}r/R_{i}\right) \\
                                                                                 {\rm i}\beta_{qi}\vec{\sigma}\cdot\hat{r}j_{1}\left(x_{qi}r/R_{i}\right) \\
                                                                               \end{array}
                                                                             \right)\frac{\chi_{q}}{\sqrt{4\pi}},
\end{equation}
where
\begin{eqnarray}\epsilon_{qi\pm}\!\!\!\!&=&\!\!\!\!\Omega_{qi}{\pm}R_{i}\left(g_{\omega}^{q}\omega_{0}
+g_{\rho}^{q}I_{3i}\rho_{03}+g_{\phi}^{q}\phi_{0}\right),\label{3}\\
\beta_{qi}\!\!\!\!&=&\!\!\!\!\sqrt{\frac{\Omega_{qi}-R_{i}m_{q}^{*}}{\Omega_{qi}+R_{i}m_{q}^{*}}},\\
\Omega_{qi}\!\!\!\!&=&\!\!\!\!\sqrt{x_{qi}^{2}+\left(R_{i}m_{q}^{*}\right)^{2}},\end{eqnarray}
with
\begin{equation}m_{q}^{*}=m_q-g_{\sigma}^{q}\sigma-g_{\sigma^{*}}^{q}\sigma^{*},\end{equation}
the effective mass of quark with flavor $q$; $R_i$ is the bag radius
of hadron species $i$; $I_{3i}$ is the isospin projection for the
hadron species $i$; $x_{qi}$ is the dimensionless quark momentum and
it can be determined from the boundary condition on the bag surface
by the eigenvalue equation
\begin{equation}j_{0}(x_{qi})=\beta_{qi}j_{1}(x_{qi}).\end{equation} In Eq. (\ref{3}), $+$
sign is for quarks and $-$ sign is for antiquarks.

The MIT bag energy is given as
\begin{equation}E^{\rm bag}_i=\sum_{q}n_{q}\frac{\Omega_{qi}}{R_{i}}-\frac{Z_{i}}{R_{i}}
+\frac43\pi R_{i}^{3}B_{i}(\sigma,\
\sigma^{*})\label{6},\end{equation} where $n_{q}$ is the number of
the constituent quarks (antiquarks) $q$ inside the bag; $Z_{i}$ is
the zero-point motion parameter of the MIT bag and $B_{i}$ is the
bag constant for the hadron $i$. In the MQMC model, the bag constant
is affected by the medium effect, and we adopte the following
directly coupling form\cite{zj}:
\begin{equation}B_{i}(\sigma,\ \sigma^*)
=B_0\exp\left[-\frac4{M_i}\left(g_{\sigma}^{{\rm bag},i}\sigma+g_{\sigma^{*}}^{{\rm bag},i}\sigma^{*}\right)\right],
\end{equation}
with $M_{i}$ is the vacuum mass of the bag. After the corrections of
spurious center of mass motion, the effective mass of a bag is given
by\cite{f}
\begin{equation}M_{i}^{*}=\sqrt{{E^{\rm bag}_i}^{2}-\langle p_{\rm
c.m.}^{2}\rangle _{i}}\label{8},\end{equation} with
\begin{equation}\langle p_{\rm c.m.}^{2}\rangle _{i}=\sum_{q}n_{q}^i\left(x_{qi}/R_{i}\right)^{2},\end{equation}
in which $n_q^i$ is the number of constituent quark(antiquark) $q$
in hadron $i$. And the radius $R_i$ of the bag is determined by
minimizing the effective mass, which gives
\begin{equation}\frac{\partial M_{i}^{*}}{\partial R_{i}}=0.\label{10}\end{equation}

Assume hadronic matter to consist of the members of the SU(3) baryon
octet and the kaon doublet. Baryons interact via ($\sigma,\ \omega,\
\rho,\ \sigma^*,\ \phi$) meson exchanges and antikaons are treated
in the same footing. Then the total Lagrangian density of the
hadronic matter in the MQMC model can be written as
\begin{eqnarray}
\mathcal{L}_{\rm
MQMC}=\!\!\!\!\!\!\!\!\!&&\sum_{B}\bar{\Psi}_{B}\left[{\rm
i}\gamma_\mu\partial^\mu-M_{B}^{*}-\left(g_{\omega}^{B}\omega_\mu\gamma^\mu
\right.\right.\nonumber\\
&&\left.\left.+g_{\rho}^{B}\frac{\vec{\tau}_B}{2}\cdot{\vec{\rho}}_\mu\gamma^\mu
+g_{\phi}^{B}\phi_\mu\gamma^\mu\right)\right]\Psi_{B}
\nonumber\\&&+\frac12\left(\partial_\mu\sigma\partial^\mu\sigma
+\partial_\mu\sigma^*\partial^\mu\sigma^*\right)\nonumber\\
&&-\frac{1}{2}\left(m_{\sigma}^{2}\sigma^{2}+m_{\sigma^{*}}^{2}\sigma^{*2}-m_{\omega}^{2}\omega_\mu\omega^\mu\right.\nonumber\\
&&\left.-m_{\rho}^{2}{\vec{\rho}}_\mu\cdot{\vec{\rho}}^\mu-m_{\phi}^{2}\phi_\mu\phi^\mu\right)\nonumber\\
&&-\frac14\left(W_{\mu\nu}W^{\mu\nu}
+\vec{G}_{\mu\nu}\cdot\vec{G}^{\mu\nu}+F_{\mu\nu}F^{\mu\nu}\right)\nonumber\\&&+\sum_{l}\bar{\Psi}_{l}\left({\rm
i}\gamma_{\mu}\partial^{\mu}-m_{l}\right)\Psi_{l}\nonumber\\&&+\mathcal
{D}_{\mu}^{*}K^{*}\mathcal {D}^{\mu}K-{M^*_{\rm
K}}^2K^{*}K,\label{13}
\end{eqnarray}
where the summation on $B$ is over the octet of baryons (p, n,
$\Lambda$, $\Sigma^{+}$, $\Sigma^{0}$, $\Sigma^{-}$, $\Xi^{0}$,
$\Xi^{-}$), $l\equiv(e^{-},\ \mu^{-}$) and the isospin doublet for
the antikaons is denoted by $K^*\equiv(K^-,\ \bar{K}^0)$,
$W_{\mu\nu}=\partial_\mu\omega_\nu-\partial_\nu\omega_\mu$,
$\vec{G}_{\mu\nu}=\partial_\mu\vec{\rho}_\nu-\partial_\nu\vec{\rho}_\mu$,
$F_{\mu\nu}=\partial_\mu\phi_\nu-\partial_\nu\phi_\mu$,
$\displaystyle\mathcal {D}_{\mu}=\partial_{\mu}+{\rm
i}g_{\omega}^{\rm K}\omega_{\mu}+{\rm i}g_{\rho}^{\rm
K}\frac{\vec{\tau}_{\rm K}}{2}\cdot\vec{\rho}_{\mu}+{\rm
i}g_{\phi}^{\rm K}\phi_{\mu}$. The form of the lagrangian is similar
to the usual relativistic mean field Lagrangian\cite{pb,gs}, except
that the effective mass is pre-determined by Eq. (\ref{8}). The
dispersion relation for $K^-$ can be easily derived from the
equation of motion, it takes
\begin{equation}\omega_{{\rm K}^{-}}=M_{\rm
K}^{*}-\left(g_{\omega}^{\rm K}\omega_{0}+g_{\rho}^{\rm
K}I_{3K}\rho_{03}+g_{\phi}^{\rm K}\phi_{0}\right).\end{equation}For
the sake of simplicity, we ignore $\bar{K}^0$ in the present work,
and include $K^-$ field only because it is the most possible one to
be $s$ wave condensation ($\vec{k}_{\rm K}=0$) in dense neutron star
matter\cite{pb}.

From Eqs. (\ref{13}) and (\ref{8}), we can derive the equations of
motion for the meson fields in uniform static matter:
\begin{eqnarray}m_{\sigma}^{2}\sigma\!\!\!\!&=&\!\!\!\!-\sum_{B}\frac{2
{J}_{B}+1}{2\pi^{2}}\int_{0}^{k_{B}}k^{2}{\rm
d}k\frac{M_{B}^{*}}{\sqrt{k^{2}+M_{B}^{*2}}}\frac{\partial
M_{B}^{*}}{\partial\sigma}\nonumber\\&&\!\!\!\!-\frac{\partial
M_{\rm K}^{*}}{\partial
\sigma}\rho_{\rm K},\label{12}\\
m_{\sigma^{*}}^{2}\sigma^{*}\!\!\!\!&=&\!\!\!\!-\sum_{B}\frac{2
{J}_{B}+1}{2\pi^{2}}\int_{0}^{k_{B}}k^{2}{\rm
d}k\frac{M_{B}^{*}}{\sqrt{k^{2}+M_{B}^{*2}}}\frac{\partial
M_{B}^{*}}{\partial\sigma^{*}}\nonumber\\&&\!\!\!\!-\frac{\partial
M_{\rm K}^{*}}{\partial
\sigma^{*}}\rho_{\rm K},\\
m_{\omega}^{2}\omega_{0}\!\!\!\!&=&\!\!\!\!\sum_{B}g_{\omega}^{B}\left(2
{J}_{B}+1\right)k_{B}^{3}/\left(6\pi^{2}\right)-g_{\omega}^{\rm K}\rho_{\rm K},\\
m_{\rho}^{2}\rho_{03}\!\!\!\!&=&\!\!\!\!\sum_{B}g_{\rho}^{B}I_{B3}\left(2
{J}_{B}+1\right)k_{B}^{3}/\left(6\pi^{2}\right)-g_{\rho}^{\rm K}\rho_{\rm K},\\
m_{\phi}^{2}\phi_{0}\!\!\!\!&=&\!\!\!\!\sum_{B}g_{\phi}^{B}\left(2
{J}_{B}+1\right)k_{B}^{3}/\left(6\pi^{2}\right)-g_{\phi}^{\rm
K}\rho_{\rm K},\label{16}
\end{eqnarray} where $J_B$ and $k_B$ are the spin projection and the fermi momentum for baryon $B$, respectively.
Using$$\frac{\partial{M_i^{*}}}{\partial\sigma}=\left.\frac{\partial{M_i^{*}}}{\partial\sigma}\right|_{R_{i}}
+\left.\frac{\partial{M_i^{*}}}{\partial{R_{i}}}\right|_{\sigma}\frac{\partial{R_{i}}}{\partial\sigma}$$
and Eq. (\ref{10}), we can give the differentiation of the effective
hadron (baryon and kaon) species mass with scalar field $\sigma$:
\begin{eqnarray}\frac{\partial M_{i}^*}{\partial
\sigma}\!\!\!\!&=&\!\!\!\!\frac{\displaystyle E^{\rm
bag}_i\frac{\partial E_{\rm
bag}^i}{\partial\sigma}-\frac{1}{2}\frac{\partial \langle p_{\rm
c.m.}^{2}\rangle _{i}}{\partial\sigma}}{M_{i}^*},\\\frac{\partial
E^{\rm
bag}_i}{\partial\sigma}\!\!\!\!&=&\!\!\!\!\sum_q\frac{n_{q}}{R_{i}}\frac{\partial\Omega_{qi}}{\partial\sigma}
+\frac{4}{3}\pi R_{i}^{3}\frac{\partial
B_{i}}{\partial\sigma},\\\frac{\partial \langle p_{\rm
c.m.}^{2}\rangle
_{i}}{\partial\sigma}\!\!\!\!&=&\!\!\!\!\frac2{R_{i}^{2}}\sum_{q}n_{q}\left(\Omega_{qi}\frac{\partial\Omega_{qi}}{\partial\sigma}
+R_{i}^{2}g_{\sigma}^{q}m_{q}^{*}\right),\\
\frac{\partial\Omega_{qi}}{\partial\sigma}
\!\!\!\!&=&\!\!\!\!-R_{i}g_{\sigma}^{q}\frac{\Omega_{qi}/2+m_{q}^{*}R_{i}\left(\Omega_{qi}-1\right)}{\Omega_{qi}\left(\Omega_{qi}-1\right)+m_{q}^{*}R_{i}/2},
\end{eqnarray}
and the differentiation with respect to $\sigma^{*}$ is likewise.

Since the time scale of a star can be regarded as infinite compared
to the typical time for weak interaction, which violates the
strangeness conservation, the strangeness quantum number is
therefore not conserved. While the $\beta$ equilibrium should be
maintained. All the $\beta$ equilibrium conditions involving the
baryon octet
\begin{eqnarray*}
p+e^-\leftrightarrow n+\nu_e,&\hskip3.5mm\Lambda\leftrightarrow
n,&\\
\Sigma^++e^-\leftrightarrow
n+\nu_e,&\hskip2mm\Sigma^0\leftrightarrow
n,&\hskip2mm\Sigma^-\leftrightarrow n+e^-+\bar{\nu}_e,\\
&\hskip2mm\Xi^0\leftrightarrow n,&\hskip2mm\Xi^-\leftrightarrow
n+e^-+\bar{\nu}_e\end{eqnarray*} may be summarized by a single
generic equation:
\begin{equation}\mu_{B}=\mu_{n}-q_{B}\mu_{e},\label{21}
\end{equation}
where $\mu_B$ and $q_B$ are, respectively, the chemical potential
and electric charge of baryon $B$
with\begin{eqnarray}\mu_{B}=\sqrt{k_{B}^{2}+M_{B}^{*2}}+g_{\omega}^{B}\omega_{0}+g_{\phi}^{B}\phi_{0}
+g_{\rho}^{B}I_{3B}\rho_{03}.\end{eqnarray} From the decay modes$$
K^-\leftrightarrow e^-+\bar{\nu}_e,\hskip1cm\mu^-\leftrightarrow
e^-+\bar{\nu}_e+\nu_\mu,$$we know that when the effective energy of
$K^-$ meson, $\omega_{{\rm K}^-}$, equals to its chemical potential,
$\mu_{{\rm K}^-}$, which in turn is equal to the electrochemical
potential $\mu_e$, $K^-$ condensation is formed, i.e.
\begin{equation}\omega_{{\rm K}^{-}}=\mu_{e}=\sqrt{k_e^2+m_e^2}=\mu_{\mu}=\sqrt{k_\mu^2+m_\mu^2}.\label{22}\end{equation}
Note that the first equal sign in Eq. (\ref{22}) is only valid when
the condensation takes place. And there are two physical constraints
on the HP phase left, they are the conservation of baryon-number and
electric charge, which are
\begin{eqnarray}\rho_{\rm
HP}\!\!\!\!&=&\!\!\!\!\frac{1}{6\pi^{2}}\sum_{B}b_{B}\left(2
{J}_{B}+1\right)k_{B}^{3},\label{25}\\ \rho^{\rm ch}_{\rm
HP}\!\!\!\!&=&\!\!\!\!\frac{1}{6\pi^{2}}\sum_{B}q_{B}\left(2
{J}_{B}+1\right)k_{B}^{3}+\frac{1}{3\pi^{2}}\sum_{l}q_{l}k_{l}^{3}-\rho_{\rm
K}.\nonumber\\\end{eqnarray} The electric charge neutrality
condition for the pure HP phase is \begin{equation}\rho^{\rm
ch}_{\rm HP}=0.\end{equation}

Therefore, the energy density and pressure for the HP are:
\begin{eqnarray}{\cal E}_{\rm HP}\!\!\!\!&=&\!\!\!\!\frac{1}{2}\left(m_{\sigma}^{2}\sigma^{2}
+m_{\sigma^{*}}^{2}\sigma^{*2}+m_{\omega}^{2}\omega_{0}^{2}+m_{\rho}^{2}\rho_{03}^{2}+m_{\phi}^{2}\phi_{0}^{2}\right)
\nonumber\\
&&\!\!\!\!+m_{\rm K}^{*}\rho_{\rm
K}+\sum_{B}\frac{2{J}_{B}+1}{2\pi^{2}}\int_{0}^{k_{B}}\sqrt{k^{2}+M_{B}^{*2}}k^{2}{\rm
d}k\nonumber\\
&&\!\!\!\!+\frac{1}{\pi^{2}}\sum_{l}\int_{0}^{k_{l}}\sqrt{k^{2}+m_{l}^{2}}k^{2}{\rm
d}k,\\
{\cal P}_{\rm
HP}\!\!\!\!&=&\!\!\!\!\frac{1}{2}\left(m_{\omega}^{2}\omega_{0}^{2}+m_{\rho}^{2}\rho_{03}^{2}+m_{\phi}^{2}\phi_{0}^{2}
-m_{\sigma}^{2}\sigma^{2}-m_{\sigma^{*}}^{2}\sigma^{*2}\right)\nonumber\\
&&\!\!\!\!+\sum_{B}\frac{2{J}_{B}+1}{6\pi^{2}}\int_{0}^{k_{B}}\frac{k^{4}{\rm
d}k}{\left(k^{2}+M_{b}^{*2}\right)^{1/2}}\nonumber\\&&\!\!\!\!+\frac{1}{3\pi^{2}}\sum_{l}\int_{0}^{k_{l}}\frac{k^{4}{\rm
d}k}{\left(k^{2}+m_{l}^{2}\right)^{1/2}}.
\end{eqnarray}

We assume the quark matter phase may occur in the core of star in
the form of unpair quark matter (UQM) and the spontaneous broken
chiral symmetry is restored so the quarks take their current masses.
To describe the UQM, the MIT bag model is adopted, where the quarks
are confined in a giant bag without dynamic freedom. If the bag
constant for UQM is $B$, the energy, pressure, baryon number and
electric charge densities at zero temperature are given by
\begin{eqnarray}{\cal E}_{\rm
UQM}\!\!\!\!&=&\!\!\!\!\frac3{\pi^{2}}\sum_{q}\int_{0}^{k_{q}}\sqrt{k^{2}+m_{q}^{2}}k^{2}{\rm
d}k\nonumber\\&&\!\!\!\!+\frac{1}{\pi^{2}}\sum_{l}\int_{0}^{k_{l}}\sqrt{k^{2}+m_{l}^{2}}k^{2}{\rm
d}k+B,\\
{\cal P}_{\rm
UQM}\!\!\!\!&=&\!\!\!\!\frac1{\pi^{2}}\sum_{q}\int_{0}^{k_q}\frac{k^{4}{\rm
d}k}{\left(k^{2}+m_q^2\right)^{1/2}}\nonumber\\&&\!\!\!\!
+\frac{1}{3\pi^{2}}\sum_{l}\int_{0}^{k_{l}}\frac{k^{4}{\rm
d}k}{\left(k^{2}+m_{l}^{2}\right)^{1/2}}-B,\\
\rho_{\rm UQM}\!\!\!\!&=&\!\!\!\!\frac1{3\pi^2}\sum_qk_q^3,\\
\rho_{\rm UQM}^{\rm
ch}\!\!\!\!&=&\!\!\!\!\frac1{\pi^2}\sum_qq_qk_q^3+\frac{1}{3\pi^{2}}\sum_{l}q_{l}k_{l}^{3}
\end{eqnarray}
with $q_q$ is the electric charge for quark $q$. The exact value of
$B$ is not fixed till now, and the phase transition point from HP to
UQM depends on its value sensitively, which will be discussed later.

Chemical equilibrium among the quark flavors and the leptons is
maintained by the following weak reactions: $$d\leftrightarrow
u+e^-+\bar{\nu}_e,\hskip5mm s\leftrightarrow
u+e^-+\bar{\nu}_e,\hskip5mm s+u\leftrightarrow d+u.$$ we can get the
equilibrium condition for the pure quark matter phase
\begin{equation}\mu_{d}=\mu_{s}=\mu_{u}+\mu_{e},\end{equation}
where \begin{equation}\mu_{q}=\sqrt{m_q^2+k_q^2}\end{equation} is
the chemical potential for the quark $q$, and can be obtained by the
$\beta$ equilibrium in mixed state. For the state where HP and UQM
coexist, i.e. the mixed phase, the quark chemical potentials for a
system in chemical equilibrium are related to those for baryon and
electron by\cite{g2}
\begin{eqnarray}\mu_{u}\!\!\!\!&=&\!\!\!\!\frac{1}{3}\mu_{n}-\frac{2}{3}\mu_{e},\label{32}\\
\mu_{d}\!\!\!\!&=&\!\!\!\!\mu_s=\frac{1}{3}\mu_{n}+\frac{1}{3}\mu_{e}.\label{33}\end{eqnarray}
Global electric charge neutrality condition must be satisfied and
the Gibbs construction requires that the pressures of two phases
should be equal at zero temperature. If the volume fraction of UQM
phase is $\chi$, then coexisting conditions are
\begin{eqnarray}&&\chi\rho^{\rm ch}_{\rm UQM}+\left(1-\chi\right)\rho^{\rm ch}_{\rm HP}=0,\\
&&{\cal P}_{\rm UQM}={\cal P}_{\rm HP}.\end{eqnarray} The energy
density and the total baryon-number density read
\begin{eqnarray}&&{\cal E}=\chi{\cal E}_{\rm UQM}+\left(1-\chi\right){\cal E}_{\rm HP},\\
&&\rho=\chi \rho_{\rm UQM}+\left(1-\chi\right)\rho_{\rm
HP}.\end{eqnarray}

We take $m_u=m_d=0,\ m_s=130$MeV\cite{e}. The bag constants and
zero-point motion parameters are calibrated to reproduce the mass
spectrum and the stable condition Eq. (\ref{10}) for the MIT-bags in
free space. Assuming the nucleon's radius to be 0.6fm, the bag
constant $B_0$ in vacuum for the nucleon can be fitted together with
the mass 939MeV. The result is $B_0^{1/4}=188.2385$MeV. In Table
\ref{tab1},
\begin{table}[h] \caption{The zero-point motion parameters
$Z_i$ and radii $R_i$ are obtained to reproduced the mass spectrum
in vacuum and Eq. (\ref{10}). And that $B_0^{1/4}=188.2385$MeV has
been fixed by the properties of nucleon. The mass spectrum adopted
here is taken from Ref\cite{e}.}\label{tab1}
\begin{center}\begin{tabular}{c|rcc}\hline\hline
  & M(MeV)  & Z  & R(fm) \\
\hline
 N & 939.0  &  2.0314 & 0.6000 \\
 $\Lambda$ & 1115.7  & 1.7913  & 0.6472 \\
 $\Sigma^{+}$ & 1189.4  & 1.6124  & 0.6731 \\
 $\Sigma^{0}$ & 1192.6  & 1.6041 & 0.6742 \\
 $\Sigma^{-}$ & 1197.4  & 1.5919 & 0.6758 \\
 $\Xi^{0}$ &  1314.8 & 1.4439  & 0.6940 \\
 $\Xi^{-}$ & 1321.3  &  1.4262 & 0.6960 \\
 $K^{-}$ & 493.7  & 1.1632  & 0.3391 \\
\hline \hline
\end{tabular}
\end{center}\end{table}
the zero-point motion parameters and bag-radii for baryons and $K^-$
are listed. And the mass spectrum for mesons transferring
interactions are listed in Table \ref{tab2}.
\begin{table}[b]
\caption{The mass spectrum (in MeV) for mesons transferring
interactions\cite{e}.}\label{tab2}
\begin{center}\begin{tabular}{c|c|c|c|c} \hline\hline
$m_{\sigma}$& $m_{\omega}$ &  $m_{\rho}$ & $m_{\sigma^{*}}$ & $m_{\phi}$ \\
\hline
550& 783  & 776  &  980 & 1020\\
\hline \hline
\end{tabular}\end{center}\end{table}

The $\sigma,\ \omega$ and $\rho$ mesons couple only to the up and
down quarks while $\sigma^*$ and $\phi$ couple to the strange quark.
We thus set
$$g_\sigma^s=g_\omega^s=g_\rho^s=g_{\sigma^*}^u=g_{\sigma^*}^d=g_\phi^u=g_\phi^d=0.$$
By assuming the SU(6) symmetry of the simple quark model we have the
relations\cite{ph}
$$\begin{array}{ll}g_\sigma^u=g_\sigma^d\equiv g_\sigma^{u,d},&
g_{\sigma^*}^s=\sqrt2g_{\sigma}^{u,d}, \\
g_{\sigma}^{i}=\left(n_{u}^{i}+n_{d}^{i}\right)g_{\sigma}^{u,d},
& g_{\sigma^{*}}^{i}=\sqrt{2}n_{s}^{i}g_{\sigma}^{u,d};\\
g_\omega^u=g_\omega^d\equiv g_\omega^{u,d},&
g_{\phi}^{s}=\sqrt2g_{\omega}^{u,d}, \nonumber\\
g_{\omega}^{i}=\left(n_{u}^{i}+n_{d}^{i}\right)g_{\omega}^{u,d},
& g_{\phi}^{i}=\sqrt{2}n_{s}^{i}g_{\omega}^{u,d};\nonumber\\
g_\rho^u=g_\rho^d\equiv g_\rho^{u,d},&
g_{\rho}^{i}=g_{\rho}^{u,d};\nonumber\\
g_{\sigma}^{{\rm{bag}},i}=\displaystyle\frac13\left(n_{u}^{i}+n_{d}^{i}\right)g_{\sigma}^{{\rm{bag}},N},&
\displaystyle
g_{\sigma^{*}}^{{\rm{bag}},i}=\frac{\sqrt{2}}{3}n_{s}^{i}g_{\sigma}^{{\rm{bag}},N}.\nonumber\end{array}$$
Then there are only four independent constants of coupling left.
Three of them are the couplings between light quarks and nonstrange
meson mean fields, i.e. $g_{\sigma}^{u,d}$, $g_{\omega}^{u,d}$ and
$g_{\rho}^{u,d}$. The last one is $g_{\sigma}^{{\rm bag},N}$
measuring the interaction between the bag constant and the scalar
$\sigma$ mean fields. We adjust them to reproduce the saturation
properties of nuclear matter: the symmetric energy index $a_{\rm
sym}$=32.5MeV, the binding energy $E_b=-$16MeV and the
compressibility $K$=289MeV at the density $\rho_{0}$=0.17fm$^{-3}$.
The four independent coupling constants are listed in Table
\ref{tab3}.
\begin{table}\caption{The four independent coupling constants}\label{tab3}
\begin{center}
\begin{tabular}[h]{c|c|c|c} \hline\hline
$g_{\sigma}^{u,d}$& $g_{\omega}^{u,d}$  &  $g_{\rho}^{u,d}$ & $g_{\sigma}^{{\rm bag},N}$ \\
\hline
0.9668& 2.6992  & 7.9327  & 6.8369\\
\hline \hline
\end{tabular}
\end{center}\end{table}

\begin{figure}[h]\centering
{\epsfig{file=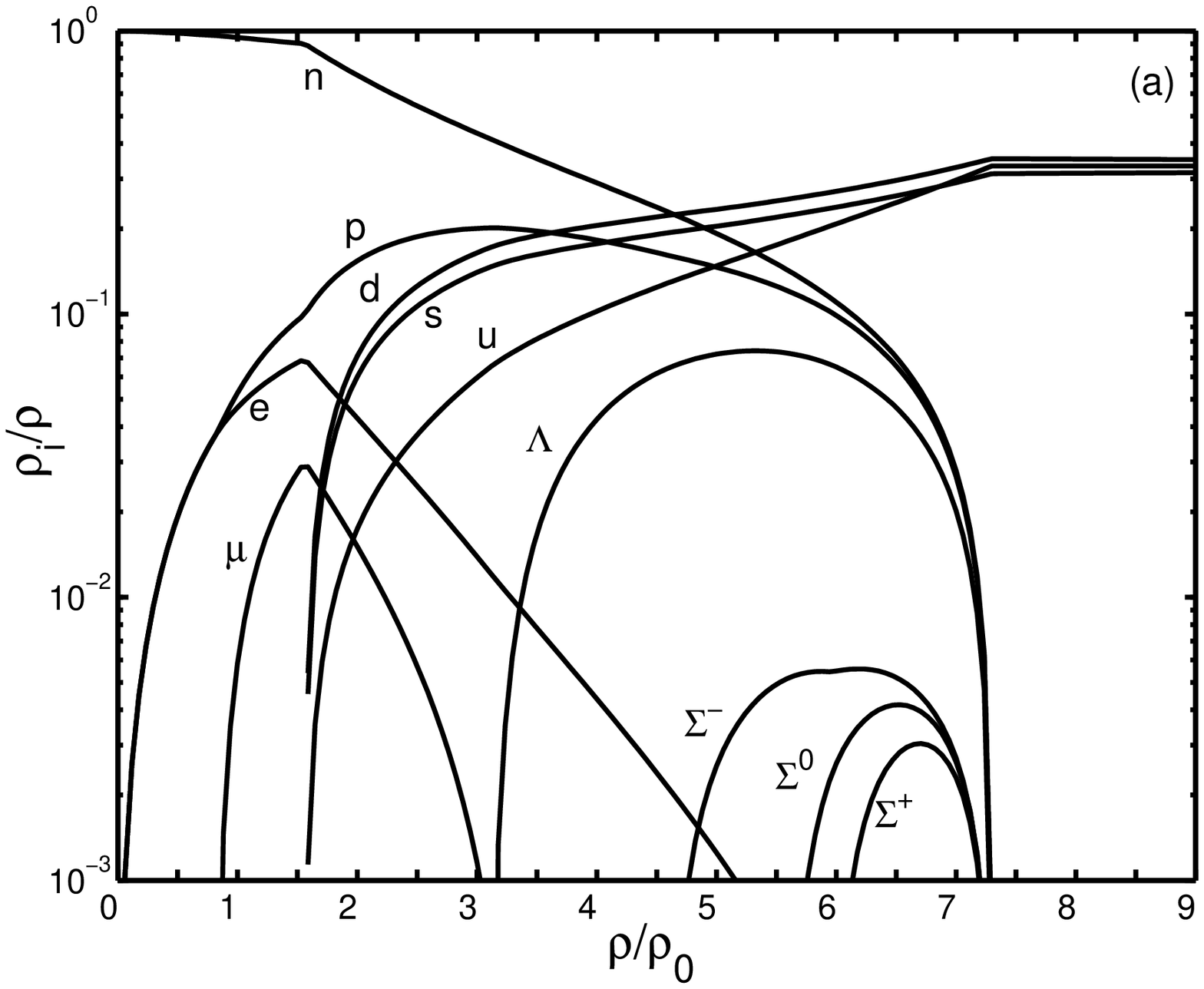,width=6cm}}
{\epsfig{file=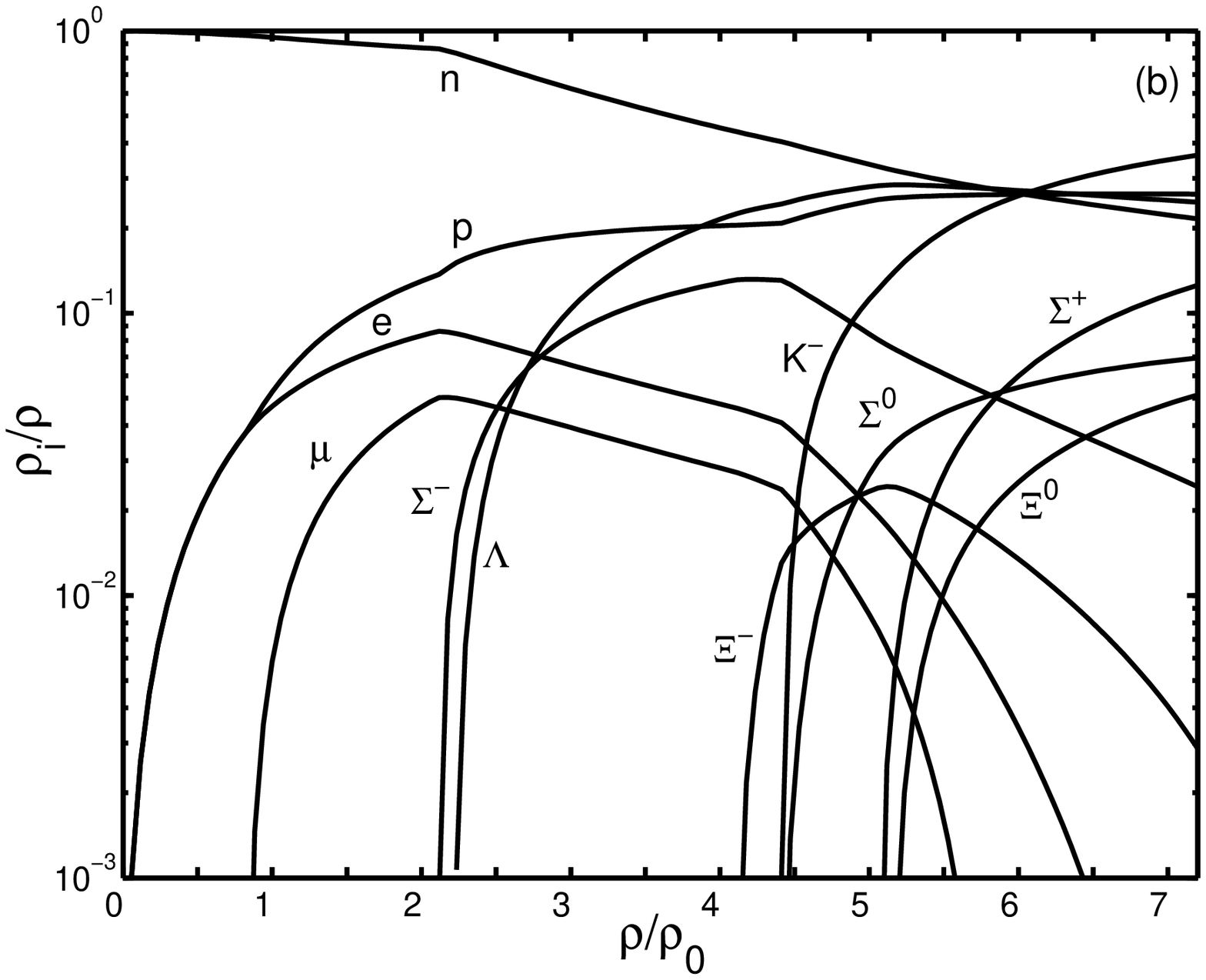,width=6cm}} \caption{The hadron, lepton
and quark population at different baryon-densities in a system
compose of (a) HP+UQM with $B^{1/4}=$180MeV, (b) pure
HP}\label{fig1}
\end{figure}

The hadron, lepton and quark population at different
baryon-densities in neutron star matter with and without UQM
respectively, are shown in Figure~\ref{fig1}.
The bag constant for UQM is fixed at $B^{1/4}$=180MeV. Figure
\ref{fig1}(a) tells us that when the density reaches 1.6$\rho_{0}$
mixed phase appears. The critical density obtained here for phase
transition from pure hadronic matter to mixed phase is similar to
those reported by other models, such as that by FST model in
Ref.\cite{gg} or the result by QMC model in Ref.\cite{p}. While in
the present model hyperons seem to appear more easily than that in
QMC model. The reason is that the effective masses of hyperons in
MQMC model are lower than that in QMC model because in the MQMC
model the bag constants of hadrons keep decreasing as the density
increases. When $\rho_{B}$=$7.8\rho_{0}$, the volume of hadronic
matter go down to zero and the pure UQM exists.

\begin{figure}[t]
\centerline{\epsfig{file=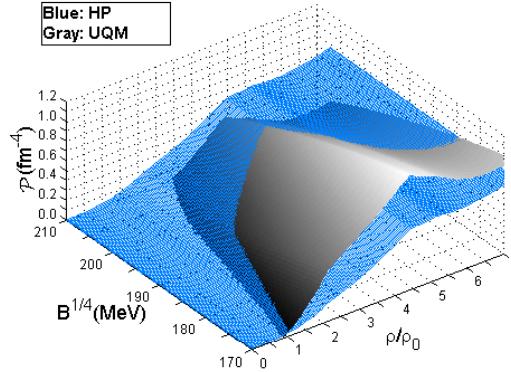,width=7cm}}
 \caption{Pressure as a function of nuclear density and bag constant.
  The light blue one is for pure HP and that gray one is
  for pure UQM using relations (\ref{32}) and (\ref{33}).}\label{fig2}
\end{figure}

In the present work, the negative charged kaon is also taken into
account, however we cannot find K$^-$ in Figure \ref{fig1}(a). To
illustrate the fact, let's look at Figure \ref{fig1}(b), which shows
the population of compositions in pure HP. It can be found that
K$^{-}$ begins to condense at a critical density of about
4.4$\rho_{0}$ which is larger than the critical density to mixed
phase. Therefore we can learn that K$^-$ condensation is suppressed
because of the deconfinement mechanism. The fact is that the
presence of UQM lowers electrochemical potential than that in pure
HP and therefore will force the critical point of condensation to a
higher density. And it is clear that the critical point has already
been forced into the region where pure UQM exists without any
hadrons. Our calculations indicate that for any choice of the bag
constant the condensed phase is suppressed once the deconfinement
phase transition is attainable, i.e. when $B^{1/4}<$202.2MeV which
we know from Figure \ref{fig2}. Our result is different from that in
Ref.\cite{gg}, where the kaon condensation appears within the mixed
phase at 9.26$\rho_{0}$ for B$^{1/4}$=185MeV. But the condensed
point is so high that the authors have to concluded that K$^{-}$
condensation would not come along with the neutron star also.

The relation between the bag constant $B$ and the critical point of
deconfinement is shown in Figure \ref{fig2}.
The light blue surface represents the pressure as a function of
$\rho$ and $B$ for HP, and the gray one is for UQM with the
conditions (\ref{32}) and (\ref{33}). The figure reveals that when
$B^{1/4}<$202.2MeV, the two surfaces can always have intersection as
the density increases, which means the system would enter the mixed
phase at the matching point. When $B^{1/4}$ is greater than about
202.2MeV, no intersection appears at all possible densities in the
interior of neutron star which means no hadron will deconfines and
the behaviors of compositions are shown in Figure~\ref{fig1}(b). For
a given $B$, pressure of quark matter phase increases firstly till
it reaches a maximum point then it drops as the density increases.
We found that the maximum is at the critical density for K$^-$
condensation and therefore with any value of $B^{1/4}<$202.2MeV
deconfiment position is always lower than that for K$^-$
condensation. Moreover, from the figure we find that the critical
density for deconfinement is sensitive to the value of $B$.

\begin{figure}[t]
  \centering
{\epsfig{file=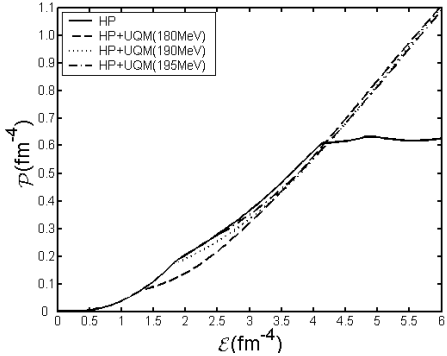,width=6cm}}
  \caption{EOSes for HP and HP+UQM with different bag constants $B$.}\label{fig3}
\end{figure}

The EOSes for pure HP and HP$+$UQM are shown in Figure~\ref{fig3}.
For pure HP, the first turning point corresponds to the appearance
of hyperons. And when the density reaches 4.4$\rho_{0}$, K$^-$
begins to condense, consequently the EOS is softened significantly.
In the system of HP$+$UQM, hyperons are forced to appear at higher
densities. At the low energy density, the EOS for HP$+$UQM is softer
than that of pure HP because of the deconfinement phase transition.
However, after the K$^-$ condensation takes place in the pure HP,
the case is contrary, which can be interpreted by two facts: First,
the abundance of hyperon is higher for pure HP at the same energy
density; Second, the $s$ wave K$^-$ condensation contributes only to
the total energy but not to pressure because of the zero momentum at
ground state.

\begin{figure}[h]
\centerline {\epsfig{file=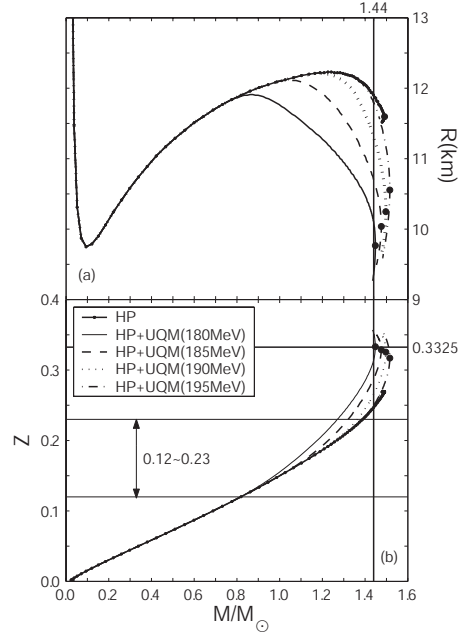,width=6cm}}
 \caption{(a)The radii versus masses for neutron stars for different EOSes.
The dots show the positions for maximum masses.The vertical line
shows the maximum mass limit by PSR 1913$+$16\cite{wt}. (b)The
gravitational redshift versus masses for neutron stars. The two
lower horizontal lines denote the observational value for the
gravitational redshift of the neutron star 1E 1207.4-5209\cite{spz}
and the one lies on 0.3325 is the lower limit for maximum
redshift\cite{c,bm}.}\label{fig4}
\end{figure}

The radius-mass relationships of static neutron star obtained by
solving the Tolman-Oppenheimer-Volkoff equations\cite{tov} are shown
in Figure \ref{fig4}(a) for different equations of state.
For all the cases studied here, the maximum masses of the stars are
found to lie between 1.45M$_{\odot}$ and 1.52M$_{\odot}$ which are
all larger than the best measured pulsar mass 1.44M$_\odot$ in the
binary pulsar PSR 1913$+$16\cite{wt}. Furthermore, for the case of
HP+UQM, a smaller bag constant gives a lower maximum mass, so the
EOSes with about B$^{1/4}<180$MeV should be ruled out.

The gravitational redshifts are plotted in Figure \ref{fig4}(b). A
redshift of z=0.35\cite{c}, with a total measurement error of order
of 5\%\cite{bm}, was inferred by identifying three sets of
redshifted transitions in the EXO0748-676 spectrum, so it imposes a
lower limit of about 0.3325 to the maximum redshift. From the
figure, we see that the EOS of pure hadron matter with condensed
K$^-$ phase is ruled out, and EOSes of HP$+$UQM with $B^{1/4}$ more
than 180MeV would be ruled out likewise, but that with 180MeV is
marginally permitted because it produces a maximum redshift of
0.3330. So the value of $B^{1/4}$ is constrained to be about 180MeV
by the combined constraint from PSR 1913$+$16 and EXO0748-676.
Sanwal et al. discussed the absorption lines from the neutron star
1E 1207.4-5209, where a redshift of 0.12$\sim$0.23 was yielded if
the observed features are identified as atomic transitions of
once-ionized helium in a strong magnetic field\cite{spz}. By the EOS
of HP+UQM with $B^{1/4}=180$MeV, we can see that the identification
corresponds to that mass M$=0.82\sim$1.27M$_\odot$ and radius
$R=11.0\sim11.9$km respectively. These values appear to be
realistic.

\begin{figure}[t]
  \centering
{\epsfig{file=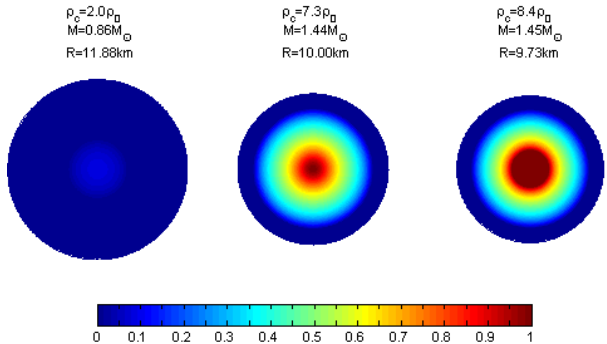,width=8cm}}
  \caption{Phase structure of hybrid stars with bag constant fixed at B$^{1/4}$=180MeV.
  The color shows the volume fraction $\chi$ of quark matter,
  and that for hadronic matter is $1-\chi$. Above the star, its properties are marked.}\label{fig5}
\end{figure}

In Figure \ref{fig5},
the phase structure in possible hybrid stars are figured out, where
$B^{1/4}$=180MeV is fixed. When $\rho_c<8.4\rho_0$, the neutron star
is stable since $\displaystyle\frac{dM}{d{\cal E}_{c}}>$0, where
${\cal E}_{c}$ is the energy density at the center. In the possible
hybrid star the density increases as going deeply into the interior
and the mixed phase exists within some critical radius. Within the
hadronic matter crust the phase transition from normal nuclear
matter into hyper-nuclear matter may occur, but K$^-$ condensation
phase is suppressed. For different hybrid stars, the volume fraction
for UQM keeps increasing, whereas the pure hadronic matter crust
becomes thinner and thinner as the central density rises. Especially
when $\rho_{c}$ goes up to about 7.3$\rho_0$, a core of pure quark
matter comes into being. And the quark core will expands further as
the $\rho_{c}$ increases. For the neutron star with maximum mass, an
evident quark matter core is presented, which is described by the
third pattern. The appearance of the pure quark core is notable.
Many other works have been carried out to study the quark matter
phase within the three-flavor NJL model, but all of them are unable
to construct a stable hybrid star with pure quark core\cite{sbb} and
only a star with mixed phase core is possible\cite{bn}.

Recently, Ransom, et. al. inferred that at least one of the stars in
Terzan 5 is more massive than 1.48, 1.68, or 1.74 M$_{\odot}$ at
99\%, 95\%, and 90\% confidence levels\cite{r}. While compared with
the limit of 1.68M$_{\odot}$, all the EOSes with exotic phases
presented here would be ruled out. Therefore, if the rather massive
star is confirmed condensed K$^{-}$ phase and deconfiment phase on
unpaired state are likely to be denied in neutron stars.

In summary, we have investigated the K$^-$ condensation and the
deconfinement phase transition in the frame of MQMC model. The model
predicts a critical density for kaon condensation in pure HP. When
UQM exists, which is only possible for $B^{1/4}<$202.2MeV, condensed
phase is suppressed. We find that only the EOS of HP+UQM with
$B^{1/4}$ about 180MeV can fit the observational mass of star PSR
1913$+$16 and the inferred redshift for EXO0748-676 at the same
time. The phase structures of possible hybrid stars with different
central densities are discussed, it is found that for EOS of
HP$+$UQM with $B^{1/4}=$180MeV a star with central density higher
than 7.3$\rho_0$ will has a pure quark core and the pure hadronic
matter star exists when $\rho_{c}<1.6\rho_0$. Between the two
densities, the star is characterized by a crust of hadronic matter
and a core of mixed phase.

The recent inferred mass of the star Terzan 5 I is also considered,
and it is found that the mass limit of 1.68M$_{\odot}$ at 95\%
confidence level makes all the EOSes presented here ruled out.
Therefore, if this rather massive star is confirmed, condensed
K$^{-}$ phase and deconfined phase in unpaired state are unlikely to
appear in neutron star by the present model and accordingly the
matter of hadrons in normal state seems to be preferred as claimed
by \"Ozel\cite{o}. Does this really mean that the ground state of
matter is composed of normal nuclear matter without exotic phases?
Actually, we should note that in the present model all the octet of
baryons are included in the HP but quarks may be deconfined within
the matter of nucleons without hyperons. Furthermore, the deconfined
quarks are in the unpaired state in the present calculation where
the quark-quark interactions are neglected, but quarks could be in
the color superconducting state as well if the the attractive
interaction in color antitriplet channel is considered. Therefore
the possibility of constructing an EOS with exotic phases which
satisfies the observational constraints could not be eliminated,
which deserves further investigations.

\noindent{\bf Acknowledgments}

The authors would like to thank Prof. Naoki Itoh for his kindly
introducing his previous work on strange quark matter. We are
grateful to Prof. Qi-Ren Zhang for valuable suggestions. Financial
support by the National Natural Science Foundation of China under
Grant Nos. 10305001, 10475002 \& 10435080 is gratefully
acknowledged.


\begin{thebibliography}{10}
\bibitem{l}J. M. Lattimer and M. Prakash, Science 304 (2004) 536.
\bibitem{k}D. B. Kaplan and A. E. Nelson, Phys. Lett. B 175 (1986) 57.
\bibitem{b}A. R. Bodmer, Phys. Rev. D 4 (1971) 1601.
\bibitem{w}E. Witten, Phys. Rev. D 30 (1984) 272.
\bibitem{i}N. Itoh, Prog. Theor. Phys. 44 (1970) 291.
\bibitem{gg}J. f. Gu, H. Guo, X. G. Li, Y. X. Liu and F. R. Xu, Phy. Rev. C 73 (2006) 055803 and references therein.
\bibitem{g1}P. A. M. Guichon, Phys. Lett. B 200 (1988) 235.
\bibitem{st}K. Saito and A. W. Thomas, Phys. Rev. C 52 (1995) 2789;

H. M\"uller, B. K. Jennings, Nucl. Phys. A 626 (1997) 966;

J. C. Caillon and J. Labarsouque, Phys. Lett. B 425 (1998) 13.
\bibitem{gsr}P. A. M. Guichon, K. Saito, E. Rodionov, and A. W. Thomas, Nucl. Phys. A 601
(1996) 349;

P. G. Blunden and G. A. Miller, Phys. Rev. C 54 (1996) 359;

H. M\"uller, Phys. Rev. C 57 (1998) 1974.
\bibitem{m}D. P. Menezes, P. K. Panda and C. Provid$\hat{\rm e}$ncia, Phy. Rev. C 72 (2005) 035802.
\bibitem{p}P. K. Panda, D. P. Menezes and C. Provid$\hat{\rm e}$ncia, Phy. Rev. C 69 (2004) 025270.
\bibitem{j}X. Jin and B. K. Jennings, Phys. Lett. B 374 (1996) 13;

X. Jin and B. K. Jennings, Phy. Rev. C 54 (1996) 1427;

X. Jin and B. K. Jennings, Phy. Rev. C 55 (1997) 1567.
\bibitem{sd}J. Schaffner, C. B. Dover, A. Gal, C. Greiner, D. J. Millener and H.
St\"ocker, Annals of Physics 235 (1994) 35.
\bibitem{ph}S. Pal, M. Hanauske, I. Zakout, H. St\"ocker and
W. Greiner, Phy. Rev. C 60 (1999) 015802.
\bibitem{zg}I. Zakout, W. Greiner, H. R. Jaqaman, Nucl. Phys. A 759 (2005) 201.
\bibitem{zj}I. Zakout, H. R. Jaqaman and W. Greiner, J. Phys. G: Nucl. Part.
Phys. 27 (2001) 1939.
\bibitem{f}S. Fleck, W. Bentz, K. Shimizu and K. Yazaki, Nucl. Phys. A 510 (1990) 731.
\bibitem{pb}S. Pal, Debades Bandyopadhyay, W. Greiner, Nucl. Phys. A 674 (2000) 553.
\bibitem{gs}N. K. Glendenning and J\"urgen Schaffner-Bielich, Phy. Rev. C 60 (1999) 025803.
\bibitem{g2}N. K. Glendenning, Phy. Rev. D 46 (1992) 1274.
\bibitem{e}Particle Data Group, S. Eidelman, K. G. Hayes, K. A. Olive, M. Aguilar-Benitez, C. Amsler, D. Asner,
K. S. Babu, R. M. Barnett and J. Beringer, et al., Phys. Lett. B 592
(2004) 1.
\bibitem{tov}R. C. Tolman, Phys. Rev. 55 (1939) 364;

J. R. Oppenheimer, G. Volkoff, Phys. Rev. 55 (1939) 374.
\bibitem{spz}D. Sanwal, G. G. Pavlov, V. E. Zavlin and M. A. Teter, Astrophys. J. 574 (2002) L61.
\bibitem{wt}J. M. Weisberg and J. H. Taylor, {\sl Radio Pulsars} ed M. Bailes,
D. J. Nice and S. Thorsett (San Francisco: Astronomical Society of
the Pacific) 2003;

S. E. Thorsett and D. Chakrabarty, Astrophys. J. 512 (1999) 288.
\bibitem{c}J. Cottam, F. Paerels and M. Mendez, Nature 420 (2002)
51.
\bibitem{bm}S. Bhattacharyya, M. C. Miller and F. K. Lamb, Astrophys. J.
644 (2006) 1085.
\bibitem{sbb}K. Schertler, S. Leupold and J. Schaffner-Bielich, Phys. Rev. C 60
(1999) 025801;

M. Baldo, M. Buballab and G. F. Burgioa, et al., Phys. Lett. B 562
(2003) 153;

M. Baldo, G. F. Burgio, P. Castorina, S. Plumari and D.
Zappal$\grave{\rm a}$, Phys. Rev. C 75 (2007) 035804.
\bibitem{bn}M. Buballa, F. Neumann, M. Oertel and I. Shovkovy, Phys. Lett. B 595 (2004) 36.
\bibitem{r}S. M. Ransom, Jason W. T. Hessels and I. H. Stairs, et al., Science 307 (2005) 892.
\bibitem{o}F. \"Ozel, Nature 441 (2006) 1115.
\end{thebibliography}
\end{document}